# Double-superradiant cathodoluminescence

**Alexey Gorlach[†1], Ori Reinhardt[†1], Andrea Pizzi[†2], Ron Ruimy[1], Gefen Baranes Spitzer[1], Nicholas Rivera[3,4] and Ido Kaminer[1]**

[1] Department of Electrical Engineering and Solid State Institute, Technion - Israel Institute of Technology, 32000 Haifa, Israel

[2] Cavendish Laboratory, University of Cambridge, Cambridge CB3 0HE, United Kingdom

[3] Department of Physics, Harvard University, Cambridge, MA 02138, USA

[4] Department of Physics, Massachusetts Institute of Technology, Cambridge, MA 02139, USA

[†] *equal contributors*

**There exist two families of superradiance phenomena: relying on correlated emission by an ensemble of atoms or by free electrons. Here we investigate emission from an ensemble of atoms driven by coherently shaped electrons. This interaction creates superradiance emerging from both the atoms' and the electrons' coherence – with emission intensity that scales quadratically in both the number of atoms and number of electrons. This phenomenon enables electrons to become novel probes of quantum correlations in matter, with high temporal and spatial resolution.**

## Introduction

Electron microscopy is ubiquitous in imaging and spectroscopy of matter [1–3]. Recent advances propose using temporal [4–6] and spatial electron shaping [7–9] to affect electron-matter interactions, such as cathodoluminescence, which is the emission of light by free electrons impinging on matter [10–16]. Cathodoluminescence is central in modern electron microscopy, providing detailed information on optical properties of samples. The idea that cathodoluminescence can be enhanced by shaped electrons has received a recent burst of interest [17–26].

At the core of our understanding of enhanced cathodoluminescence by shaped electrons is the coherent interaction between a single two-level system and shaped free electrons. This interaction has been first investigated using a semiclassical theory [17,18], proposing coherent control of the two-level system and enhancement of cathodoluminescence [17]. Soon after, the fully-quantum regime of this interaction was revealed [19–21,25], leading to new concepts for imaging the coherent state of matter [20], resolving the wavefunction of the emitting free electrons [23], and performing homodyne-type measurements [24]. In all these works, the enhanced electron-matter interaction scales quadratically in the number of electrons, similar to the phenomenon of electron superradiance [27–33]. Electron superradiance provides a source of coherent radiation in celebrated examples such as free-electron lasers [32], klystrons [31], synchrotrons [27–29], and Smith-Purcell sources [30].

Atomic Dicke superradiance, which differs from electron superradiance, is the collective radiation by multiple excited atoms that exhibits faster rates of spontaneous emission than by individual atoms [34,35]. Throughout the paper, we use the word "atom" to refer to any emitter from which superradiance can occur, including actual atoms, molecules, and quantum dots. Atomic superradiance plays an important role in quantum optics [36,37], relativity [38], astrophysics [39], and condensed matter [40]. However, atomic superradiance has never been considered in the context of cathodoluminescence. It remains unknown how the superradiant atomic system interacts with free electrons and whether atomic correlations can additionally enhance cathodoluminescence.

Here we consider the interaction of shaped electrons with a many-body system, finding an interplay of atomic and electron superradiance, which results in a drastic enhancement of cathodoluminescence. We present a quantum theory of the interaction between an ensemble of $N_{\mathrm{a}}$ atoms and $N_{\mathrm{e}}$ free electrons, capturing the effect of free-electron-induced superradiance (Fig. 1). Depending on whether the atoms are in the superradiant state or not, and whether the free electrons are shaped or not, we identify four possible regimes of superradiance, with emission intensity $I$ scaling as $N_{\mathrm{a}}N_{\mathrm{e}}$, $N_{\mathrm{a}}N_{\mathrm{e}}^2$, $N_{\mathrm{a}}^2N_{\mathrm{e}}$, or $N_{\mathrm{a}}^2N_{\mathrm{e}}^2$, (Fig. 1a-d). The most interesting case is the last one, occurring when the atoms emit coherently after excitation from shaped electrons, with an interplay of both electron and atomic superradiant effects. The superradiant phenomena happening here can be detected not only in the emission of light but also in the energy spectrum of the electrons, as discussed further in [41]. The preliminary results that led to our work here were first presented in [42].

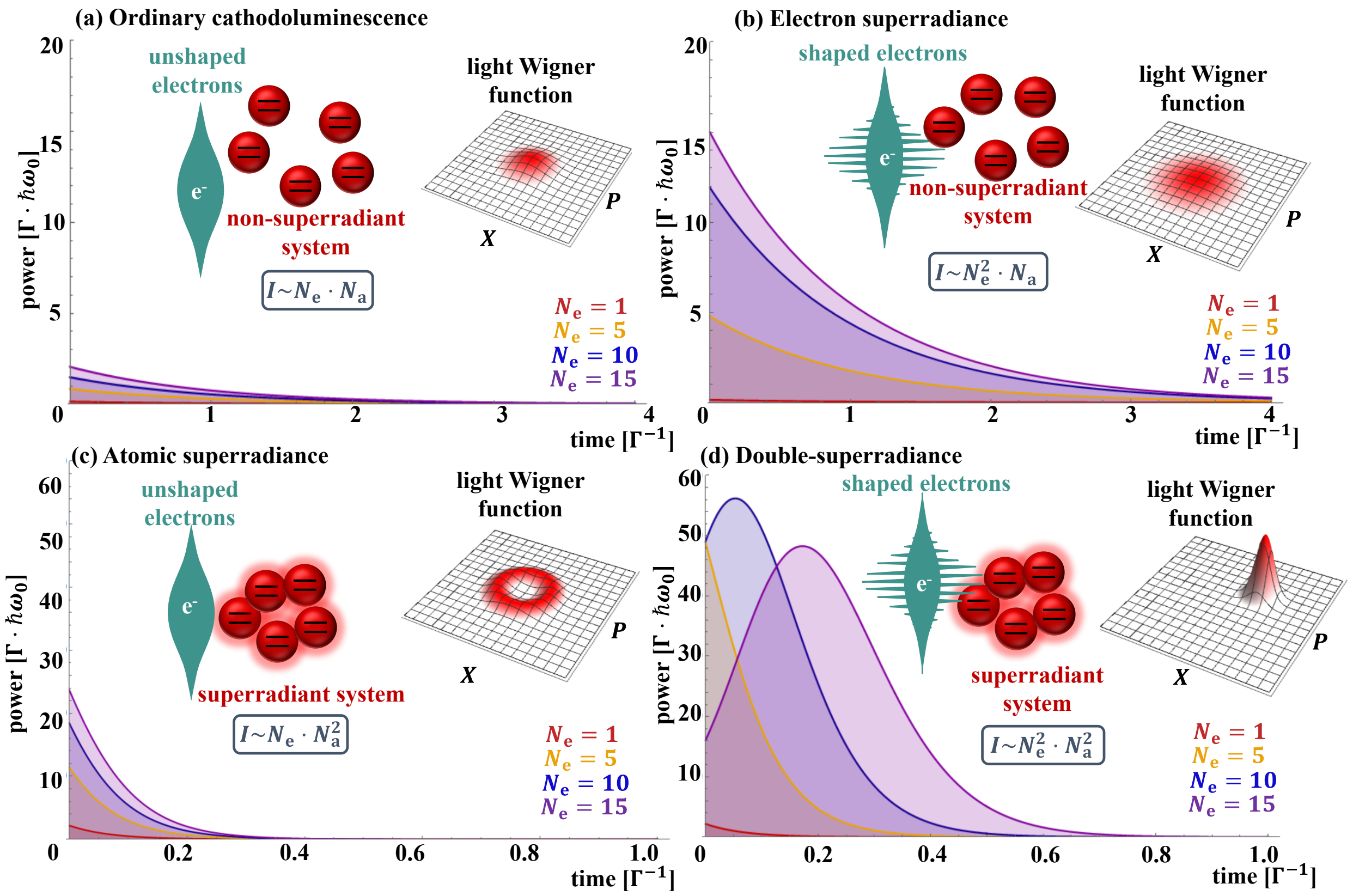

**Figure 1: Double-superradiant cathodoluminescence. (a)-(b)** Emission intensity of the non-superradiant atomic system after the interaction with unshaped ($\sigma = 0.0$) and comb electrons ($\sigma = 3.0$), respectively. The emission intensity from a single electron ($N_e = 1$) does not depend on the electron shape. **(c)-(d)** Emission intensity of the superradiant atomic system after the interaction with unshaped ($\sigma = 0.0$) and comb electrons ($\sigma = 3.0$), respectively. The plots assume $g = 0.1$ and $N_a = 15$ (the coupling constant $g$ is defined below after Eq. (3)).

Let us discuss how the double-superradiant regime (Fig. 1d) arises. The shaped electrons interact with the atomic system coherently such that the excitations they induce accumulate in a coherent manner [17,20,21]. Thus, the total energy of the atomic system after the excitation by $N_e$ electrons will be proportional to $N_e^2$. If we assume that the electrons have the same interaction with all the atoms, then the excited state is described by Dicke superradiant states [34]. The resulting emission follows the Dicke theory [35], with peak emission intensity proportional to $N_a^2$. Since the emission intensity depends on the initial excitation, we get that both effects of atomic and electron superradiance are combined so that $I$ scales as $N_a^2 N_e^2$.

The shaping of electrons discussed in this work is possible thanks to the recent advances in ultrafast electron microscopy experiments, and specifically the advent of photon-induced nearfield electron microscopy (PINEM) [2,4,5,43–48].

## Interaction between free electrons and a many-body atomic system

In this section, we develop the theory of the interaction between free electrons and an ensemble of atoms. We model each atom as a two-level system, which is the standard approach for the description of both the electron-atom interaction [17,20–22] and superradiance [34]. We should note that the electron can undergo multiple competing transitions including excitations of core levels in the atoms and emission of Bremsstrahlung [49] (elaborated in [20]). These competing transitions occur with smaller probabilities for shaped electrons as the shaping makes the interaction resonant for the specific atomic transition that matches the shaping frequency [17]. Furthermore, when looking at the cathodoluminescence signal, neglecting the competing transitions is even more relevant since the two-level emission is further enhanced by superradiance [34]. Achieving the conditions of superradiance requires having the distance between atoms smaller than the wavelength of the emitted light and electron-atom interaction length.

We denote the ground and excited states of every single atom by $|\mathrm{g}\rangle$ and $|\mathrm{e}\rangle$. The Dicke symmetric states are

$$\begin{cases} |0\rangle \equiv |\mathrm{ggg} \ldots \mathrm{gg}\rangle, \\ |1\rangle \equiv \sqrt{1/N_\mathrm{a}}(|\mathrm{egg} \ldots \mathrm{gg}\rangle + |\mathrm{geg} \ldots \mathrm{gg}\rangle + \cdots + |\mathrm{ggg} \ldots \mathrm{ge}\rangle), \\ \ldots \\ |N_\mathrm{a}\rangle \equiv |\mathrm{eee} \ldots \mathrm{ee}\rangle. \end{cases} \tag{1}$$

The Hamiltonian of a single electron interacting with the superradiant system (in analogy to [20]) reads

$$H = -i\hbar v \partial_z + \frac{\hbar\omega_0}{2} S_z + \frac{e}{4\pi\varepsilon_0} \cdot \frac{(\boldsymbol{d}_\perp \cdot \boldsymbol{r}_\perp + d_\parallel z)S_+ + (\boldsymbol{d}_\perp^* \cdot \boldsymbol{r}_\perp + d_\parallel^* z)S_-}{(z^2 + r_\perp^2)^{3/2}}. \tag{2}$$

The first two terms describe the Hamiltonians of the free electron and the superradiant system, respectively. $S_\pm = \sum_i (\sigma_i^x \pm i\sigma_i^y)$, $S_z = \sum_i \sigma_i^z$ are the symmetric operators, with $\sigma_i^{x,y}$ the Pauli operators of the $i^{\text{th}}$ atom. $v$ is the average velocity of the electron and $\hbar\omega_0$ is the transition energy of the two-level atoms. The third term in Eq. (2) describes the interaction between the electron and the atoms with transition dipole moment $\boldsymbol{d}$, having the perpendicular $\boldsymbol{d}_\perp$ and parallel $d_\parallel$ components relative to the electron trajectory. The distance between the electron and the atoms is $r_\perp$; the elementary charge is $e$ and vacuum permittivity is $\varepsilon_0$. Eq. (2) is valid under the paraxial-electron approximation, which is justified in all relevant experiments (e.g., [1–3]), considering that typical electron energy values (~100 keV) are much larger than typical transition energies ($\hbar\omega_0 \sim 1$ eV).

Under the above conditions, we use the Magnus expansion [50] to describe the scattering matrix of the interaction between one electron and the many-body atomic system:

$$U = e^{-i(gbS_+ + g^* b^\dagger S_-)}, \tag{3}$$

where $b$ and $b^\dagger$ are the electron energy shift operators, reducing and increasing the electron energy by quanta of $\hbar\omega_0$. $g$ is a dimensionless coupling coefficient that according to [17,20] equals:

$$g = \frac{ed_\perp \omega_0}{2\pi\varepsilon_0 \hbar v^2} K_1\left(\frac{\omega_0 r_\perp}{v}\right) + i\frac{ed_z \omega_0}{2\pi\varepsilon_0 \hbar v^2} K_0\left(\frac{\omega_0 r_\perp}{v}\right).$$

The Hamiltonian in Eq. (2) is valid when $|g| \ll 1$, which describes all known practical situations.

We express the scattering matrix (Eq. (3)) in the symmetric states basis:

$$\langle m|U|n\rangle \equiv b^{n-m} U_{mn}, \tag{4}$$

$$U_{mn} = \sqrt{m!\,n!\,(N-n)!\,(N-m)!} \sum_{k=0}^{n} \frac{(-1)^k (\cos|g|)^{N-m+n-2k} (i\sin|g|)^{m-n} (\sin|g|)^{2k}}{k!\,(n-k)!\,(m-n+k)!\,(N-m-k)!}.$$

The post-interaction atomic density matrix after tracing out the electron is

$$\rho_{\mathrm{f}}^{kl} = \sum_{m,n} \langle \psi_{\mathrm{e}}| b^{(m-k)-(n-l)} |\psi_{\mathrm{e}}\rangle U_{km} \rho_{\mathrm{i}}^{mn} U_{nl}^\dagger, \tag{5}$$

where $|\psi_{\mathrm{e}}\rangle$ is the wavefunction of the electron before the interaction.

Eq. (5) provides the density matrix of the atoms after their interaction with a single electron of any *arbitrary* wavefunction $|\psi_{\mathrm{e}}\rangle$. To show double-superradiance (Fig. 1c,d), we apply Eq. (5) $N_{\mathrm{e}}$ times sequentially. Each iteration obtains a new atomic density matrix that serves as the initial condition for the interaction with the next electron. This procedure is correct when the electron pulse duration and the electron-atoms interaction duration are much shorter than the superradiance time. To justify this assumption, consider typical pulses of $N_{\mathrm{e}} = 10$ electrons in a duration of $\tau = 1$ ps, which is much shorter than even the faster superradiant times (e.g., $t \sim \Gamma^{-1} N_{\mathrm{a}}^{-1} \sim 67$ ps, estimated for $N_{\mathrm{a}} = 15$ atoms with an intrinsic lifetime of $\Gamma^{-1} \sim 1$ ns).

## Effect of the electron shape on the many-body atomic system

We now show how the electrons' shape impacts the atomic state. The electrons' shaping is defined by the moments $\langle b^n \rangle$ (i.e., bunching factor [51]). The electron that has all $|\langle b^n \rangle| = 1$ has the most coherent interaction with atomic systems. To realize $|\langle b^n \rangle| \approx 1,$ we need electrons of wide coherent energy uncertainty, or equivalently, comb electrons [52], made from a sequence of coherent energy peaks [53–55]. The phase of the wavefunction should scale linearly in energy (as created using established techniques [53,54]). Thus, we consider electron states such as:

$$|\psi_e\rangle = \frac{1}{\sqrt{\text{norm}}} \sum_n e^{i\phi n} e^{-\frac{n^2}{2\sigma^2}} |E_n\rangle, \tag{6}$$

where $\phi \in [0,2\pi]$ is an arbitrary phase, $\sigma$ is the dimensionless comb bandwidth, and $|E_n\rangle = |E_0 + n\hbar\omega_0\rangle$ are the monoenergetic electron states (in practice, requiring energy width $< \hbar\omega_0$, as in many electron microscopes [1]) with normalization $\langle E_m | E_n \rangle = \delta_{mn}$. The initial electrons in conventional electron microscopes , termed unshaped electrons, have $\sigma < 1$, can be shaped into comb electrons with $\sigma{\sim}3$ [53,54]. Much larger $\sigma$ values were demonstrated [52,56], but without satisfying the linear-phase requirement.

Fig. 2 compares the atomic density matrix after interacting with unshaped ($\sigma < 1$) and shaped comb electrons ($\sigma \sim 3$). The effect of such interactions is analogous to excitations by a many-photon Fock state and a coherent state of light, respectively. Thus, the electrons can emulate excitations by quantum photonic states such as Fock states. The creation of such "electron Fock states" is much simpler than the creation of Fock light states (being the initial electron state from typical electron guns [1]).

Free electrons can be also used to control the post-interaction quantum state of the atoms, and in turn, the quantum state of the light subsequently emitted from the atoms (Wigner functions shown in Fig. 1). Fig. 1a: non-superradiant atoms emit incoherent light in ordinary cathodoluminescence. The emission is thermal and does not have a definite phase because of averaging over a large number of independent atoms. Fig. 1b: the thermal emission can be enhanced by electron superradiance, but since the atoms are still independent, the light emission phase remains random. Fig. 1c: superradiant atoms emit light with Poissonian statistics and uncertain phase when driven by unshaped electrons. The uncertainty in the phase of the emission is caused by the entanglement between unshaped electrons and atoms.

Fig. 1d: in contrast with the above, only the superradiant emission triggered by shaped electrons creates coherent states of light with a fixed phase. This phase is then locked to the shaped electrons' phase. The case of a single electron interacting with a superradiant atomic system is also a particular case of Fig. 1d, as long as the electron is pre-shaped. In this case, the emission will have a certain phase, but without electron superradiance since $N_\mathrm{e} = N_e^2 = 1$. Similarly, the interaction of multiple shaped electrons with a single atom will cause the emission of a coherent state, but without atomic superradiance since $N_\mathrm{a} = N_\mathrm{a}^2 = 1$.

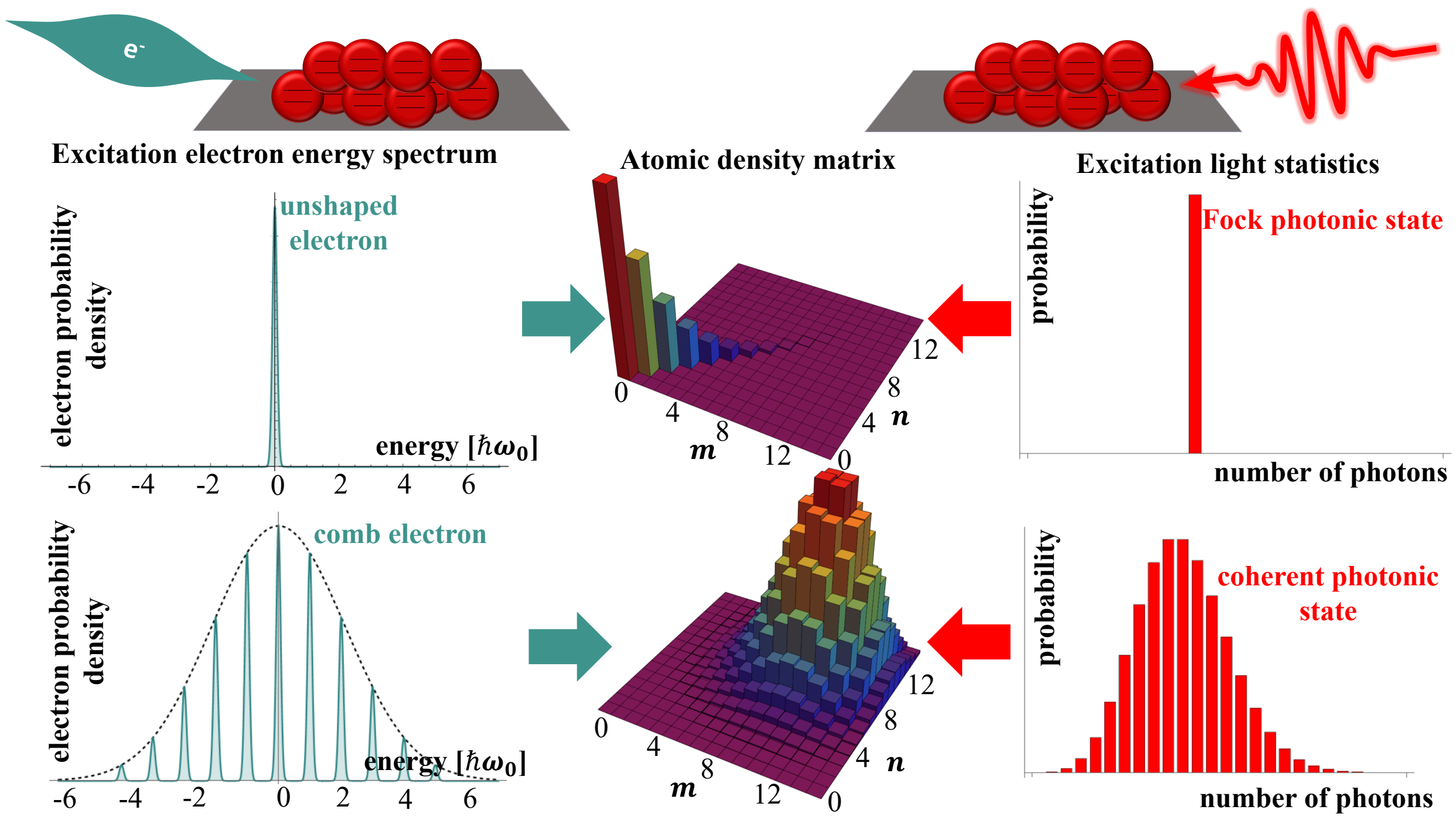

**Figure 2: Analogy between electron-driven and light-driven superradiance.** The excitations of the superradiant atomic system by a Fock and a coherent light state are analogous to the excitations by an unshaped and a comb electron respectively. This analogy is correct because unshaped free electrons and Fock states of light entangle with the atomic system, while comb electrons and coherent light do not entangle with the atomic system. In cases of entanglement, the atomic system loses its coherence after the interaction (diagonal density matrix, top row). In cases of no entanglement, the information about the phases between different atomic states remains unchanged after the interaction (off-diagonal elements in the density matrix, bottom row).

The same effects described here for a (*quantum*) comb electron wavefunction are also relevant for (*classical*) bunched electrons, whose classical temporal distribution is analogous to the wavefunction shape. The classical interaction time is similarly related to the phase $\phi$ (from Eq. (6)), determining the phase of the emitted light. Thus, the word "shaping" applies both to the quantum wavefunction and classical distribution. The influence of the electron bandwidth $\sigma$ (or bunch width) on the post-interaction atomic state is shown in Fig. 3. The probability of the atoms' excitation increases (by $N_\mathrm{e}$) for broader electron energy envelopes (Fig. 3a). A similar classical-quantum correspondence appears in electron-light interactions, which also depend on the electron energy bandwidth [57].

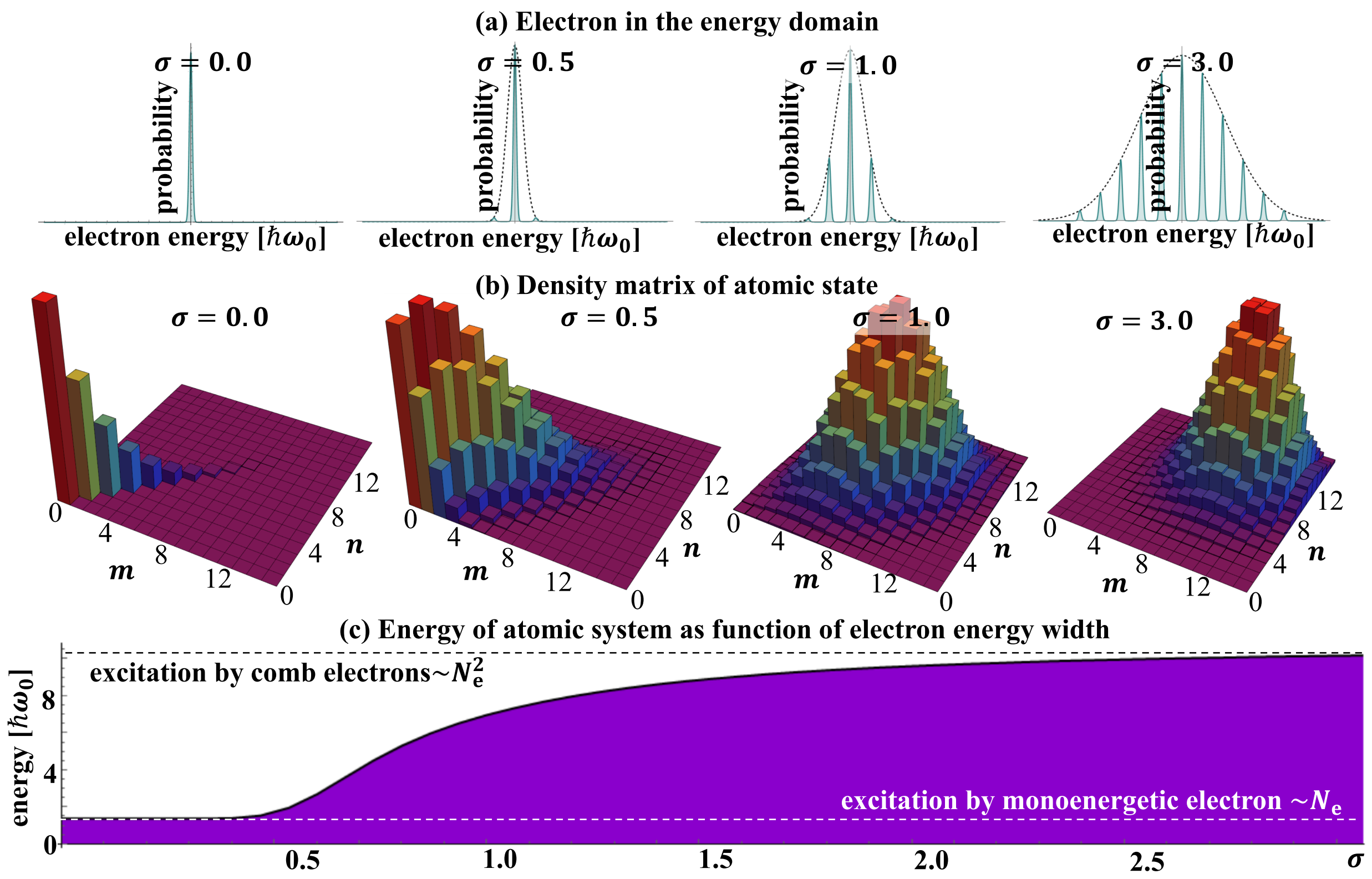

**Figure 3: The effect of electron shaping on inducing correlations between atoms. (a)** The electron energy spectrum before the interaction for different energy bandwidths $\sigma$ of the shaped electron. A wider bandwidth corresponds to stronger bunching. **(b)** The atomic density matrix after the electron interaction for different bandwidths $\sigma$. **(c)** The average atomic excitation energy following an interaction with $N_e$ electrons as a function of their energy bandwidth $\sigma$. The narrow energy bandwidth electron excitation scales as $N_e$, while the wide energy bandwidth electron excitation scales as $N_e^2$. Parameters are as in Fig. 1.

## Free-electron-induced superradiant emission

To analyze the superradiance by the atomic system, we note that the electrons (shaped or unshaped) excite only symmetric states. Due to the short atoms-electrons interaction duration (typically $< 1$ ps), we can split the dynamics into two sequential steps: (1) calculate the atomic density matrix after the interaction with $N_e$ electrons according to Eq. (5); (2) then use this density matrix as the initial condition for a conventional calculation of Dicke superradiance [34–36].

The following calculation can be used to analyze both electron-driven and light-driven superradiance. Let us call $\rho(0)$ the post-interaction atomic density matrix, as obtained from Eq. (5). The time-evolution of the atomic density matrix during the superradiance is given by [34–36]:

$$\frac{d\rho^{mn}(t)}{dt} = -\Gamma/2\big(m(N-m+1)+n(N-n+1)\big)\rho^{mn}(t) + \\ +\Gamma\sqrt{(m+1)(N-m)(n+1)(N-n)}\,\rho^{(m+1)(n+1)}(t), \quad (7)$$

where $\Gamma$ is the rate of spontaneous emission of an individual atom. The emission intensity is obtained through energy conservation as $I = -\hbar\omega_0 \sum_{m=0}^{N} m \frac{d\rho^{mm}(t)}{dt}$ (Fig. 1). Beyond the intensity, we can characterize quantum correlations of the emitted light using the model developed in [36,37], and generalized to yield the full Wigner function [58,59]. We use this development to show the Wigner function of the emitted light in Fig. 1.

## Discussion

Double-superradiance can be observed in current experimental setups. Such an observation would access the hard-to-reach coherent interaction between electrons and atoms. Previous theoretical works in this field (e.g., [17,20,21]) showed that the electron-atom interaction depends on the electrons' shape. However, this effect remained inaccessible experimentally because of the weak interaction (typical $|g|\sim 10^{-2}-10^{-3}$) in realistic systems. The intensity of emission from individual atoms is similarly weak ($I \propto |g|^2 < 10^{-4}$). Double-superradiance could resolve this challenge by enhancing the interaction by $N_\mathrm{a} \cdot N_\mathrm{e}$ times (e.g., 150 times for $N_\mathrm{a} = 15$ and $N_\mathrm{e} = 10$). Such enhancement can help surpass competing decoherence channels that so far prevented the observation of electron-atom interactions.

Double-superradiance may also help demonstrate the so far unobserved influence of electrons' shape on electron-atoms interactions. The strength of the interaction and the consequent emission depend on the shape as shown in Fig. 3. The shape also alters the quantum state of the emitted light (e.g., photon statistics), which can be extracted using quantum optical characterizations such as $g^{(2)}$ and homodyne measurements.

A proposal for the experimental implementation of double-superradiance is illustrated in Fig. 4. The electrons are pre-shaped into an energy comb using interactions with intense laser light [53,54]. These shaped electrons then interact with the atomic system, inducing emission that is collected and analyzed to extract the classical time-dependent intensity of the emitted light, for example using a streak camera. and Quantum properties of the light such as its Wigner functions (Fig. 1) can be extracted, for example using homodyne schemes that rely on a local oscillator based on the shaped electrons or on the laser that shaped them (as in [24]). Such experiments can be implemented in ultrafast (laser-driven) transmission or scanning electron microscopes – Figs. 4a,b respectively.

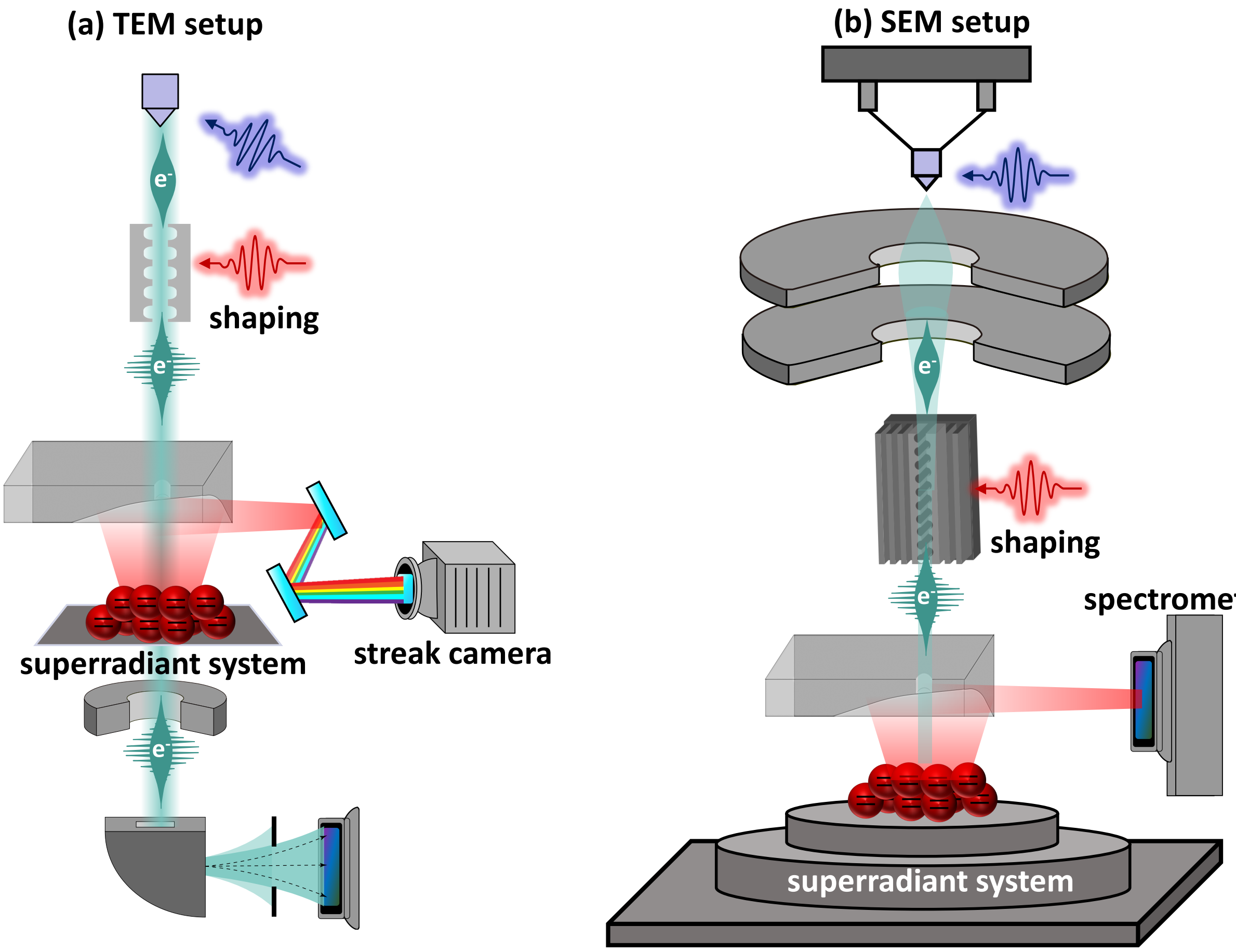

**Figure 4: Experimental schemes for measuring double-superradiant cathodoluminescence.** Schematic setups of an ultrafast **(a)** transmission or **(b)** scanning electron microscope (TEM or SEM). A pulse of electrons can be photo-induced and shaped using a laser and a coupling structure, e.g., dielectric laser accelerator structures [60,61] or a set of mirrors [52]. Each shaped electron pulse excites an ensemble of atoms. The atoms then relax through superradiant emission, which can be collected and characterized in various ways, such as a streak camera for combined spectral and temporal information. The electrons' energy spectra can be simultaneously measured for additional information, as shown recently in [62,63].

Looking forward, free electrons may become prime candidates to investigate superradiance with high spatial and temporal resolution beyond the limits enforced by conventional light excitations. The spatial resolution of light excitations is typically limited by the wavelength, i.e., hundreds of nanometers, making it impossible to selectively trigger superradiance in small regions of a target sample. In contrast, electrons enable triggering superradiance in selected small regions of the sample, enabling to image correlations between atoms. The spatial resolution of electron excitations is much better due to their much smaller de Broglie wavelength that can reach a few nanometers in scanning electron microscopes and even sub-nanometer in transmission electron microscopes.

The temporal resolution of resonant light excitations is typically limited to hundreds of femtoseconds because shorter pulses involve broader bandwidths that are mostly out-of-resonance for the intended transition. In contrast, electrons carry higher energies so they can have the broad energy bandwidth required for forming attosecond pulses [64] while maintaining a resonant excitation of the intended transition.

Altogether, we envision that using electrons instead of light could provide novel methods to control many-body quantum systems. Double-superradiant cathodoluminescence can also give access to new superradiance-related imaging capabilities, even for matter in non-equilibrium states.